\documentclass[pr,twocolumn,amsmath,nofootinbib,longbibliography,floatfix]{revtex4-1}
\usepackage{ifpdf}
\ifpdf
   \usepackage[unicode=true,colorlinks=True]{hyperref}
\fi
\usepackage{graphics}

\DeclareMathOperator{\Tr}{Tr}
\renewcommand\vec[1]{\ensuremath\boldsymbol{#1}}

\begin{document}

\title{Shape effect on the valley splitting in lead selenide nanowires}

\author{I.~D.~Avdeev}
\email{ivan-avdeev505@mail.ru}
\affiliation{Ioffe Institute, St. Petersburg 194021, Russia}

\begin{abstract}
	We study the cross-section shape and size effects on the valley splitting in PbSe nanowires within the framework of empirical $sp^3d^5s^*$ tight-binding method.
	We consider ideallized prismatic nanowires, grown along [110], with the cross-section shape varying from rectangular (terminated by $\{001\}$ and $\{110\}$ facets) to rhombic (terminated mostly by $\{111\}$ facets). 
	The valley splitting energies have the maximal value (up to hundreds meV) in rectangular nanowires, while in rhombic ones they are almost absent.
	The shape dependence is shown to be similar for a wide range cross-section sizes and different point symmetries of nanowires.
\end{abstract}

\maketitle

\section{Introduction}
Lead chalcogenide nanowires (NWs) have a wide range of possible applications. 
They can be used as circuit components~\cite{Talapin05,Kim2011}, light emitting (detecting)~\cite{Sukhovatkin09,Padilha2010} and energy harvesting~\cite{Gesuele2012,Davis15} devices. 
Throughout the past decades a big effort has been made towards experimental NW growth techniques~\cite{Lifshitz2003,Cho2005,Fardy07,Talapin2007,Koh2010,Akhtar2012}. 
They include colloidal synthesis~\cite{Talapin2007}, solution-liquid-solid~\cite{Akhtar2012} methods and even oriented attachment of lead salt nanocrystals~\cite{Cho2005}.
Nowadays it is not only possible to grow a nanowire a few nm thin~\cite{Graham2012}, but also to control its shape~\cite{Jun2006,Mokari2007,Jawaid2011}, size~\cite{Tischler2015} and growth direction~\cite{Jang2010}. 
Nonetheless, theoretical modelling of lead chalcogenide NWs is quite challanging due to the multivalley band structure and strong intervalley coupling in these systems. 
It was shown~\cite{Zunger07} that the valley splitting can exceed the excitonic exchange splitting, so the study of the valley splitting is very important to understand the fundamental electronic and optical properties of PbSe NWs.

Lead selenide is a narrow direct band gap semiconductor ($0.17$ eV~\cite{Zunger1997}) with the band extrema located at four inequivalent $L$ valleys.
In bulk crystal the ground electron and hole states are eight-fold degenerate by spin and valley degree of freedom, while in NWs this degeneracy is removed.
There are two main mechanisms that split the valley multiplets (sets of confined states originated from different valleys) in PbSe NWs: the mass anisotropy~\cite{Bartnik10} and the intervalley coupling~\cite{Delerue2004b} at the NW surface~\cite{Nestoklon06}.
The first one is readily incorporated in the $\boldsymbol{k} \cdot \boldsymbol{p}$ and splits the valley multiplets in NWs only partially~\cite{Bartnik10}. 
The second one fully removes the valley degeneracy and can be included in the $\boldsymbol{k} \cdot \boldsymbol{p}$ phenomenologically~\cite{Nestoklon06}, but for careful theoretical description it requires an atomistic approach~\cite{Zunger06,Abhijeet11} as it is very sensitive to the microscopic structure of the NW~\cite{Poddubny12,Avdeev17}.

It was shown~\cite{Avdeev17} for cylindrical PbSe NWs and spherical quantum dots~\cite{Poddubny12}, that the valley splitting depends on diameter and the point symmetry of the considered structure.
In this work we also take the shape of the NW into account. 
Following the theoretical~\cite{Argeri2011,Deringer16} and experimental~\cite{Cho2005} data, PbSe nanowires tend to have faceted structure, therefore instead of cylindrical it is more convenient to consider prismatic shape of the NWs.
For simplicity we consider idealized prismatic wires, carved out from ideal bulk PbSe crystal.
With this approach, the microscopic structure of NWs is fully determined by the cross section and the spatial orientation of the covering prism.
Even though the most natural growth direction of PbSe NWs is $[001]$~\cite{Cho2005}, we consider $[110]$ as the growth direction for the following reason: along this axis the two pairs of the $L$ valleys remain inequivalent~\cite{Abhijeet11} which allows us to explicitly evaluate the valley splitting energies for each pair and study their dependencies on the NW shape parameters.

Among many atomistic methods we chose the empirical tight binding approach in nearest neighbour approximation, proved to be useful for many cubic semiconductors~\cite{Jancu98}. 
The use of the recent parametrization~\cite{Poddubny12}, which accurately reproduces experimental effective masses in the $L$ valleys and show good agreement with experiment~\cite{Poddubny17}, allow us to study a wide range of NW sizes and shapes. 
We neglect the surface passivation, as the surface states in highly ionic crystals lie far outside the band gap~\cite{Delerue2004b} and there is no need to saturate the dangling bonds~\cite{Abhijeet11}.
Note, even though the {\it ab initio} methods, such as DFT~\cite{Wrasse11,Liu2018} and GW~\cite{Hummer07,Svane10}, are more accurate, complexity and demand in computational power make them hardly applicable to relatively large structures (more than a few hundreds of atoms).

In this paper we show, that the valley splitting drastically depends on the NW shape. 
The valley splitting is almost absent in NWs with the surface terminated mostly by $\{111\}$ facets and has the maximal value in NWs with $\{110\}$ and $\{001\}$ terminated surface.
We also notice that these are the most stable PbSe surfaces~\cite{Argeri2011,Deringer16}.
Despite the polar nature, the $\{111\}$ facets are also present in real systems, especially in colloidal solvents, where they can be passivated by ligands~\cite{Zherebetskyy2014,Liu2018}.

\section{valley coupling in nanowires}
The structure of the valley splittings is easier to analyse using the perturbative approach~\cite{Nestoklon06}.
Without the valley coupling all the $L$ valleys are independent and their energy spectra are fully determined by mass anisotropy~\cite{Bartnik10} and quantum confinement~\cite{Goupalov11}.
The latter we associate with a set of quantum numbers $q$, well established for cylindrical PbSe nanowires~\cite{Avdeev17}. 
It is important, that sets of states with a certain $q$ (valley multiplets) are well distinguishable in the tight binding~\cite{Avdeev16,Avdeev17,Pawlowski2018}, where we consider the abrupt NW boundary as a perturbation that mix the valley states.

For further discussion, we enumerate the $L$ valleys and corresponding wave vectors $\vec{k}$ by index $\mu \in \{0,1,2,3\}$ 
\begin{equation}
	\label{eq:kks}
	\begin{array}{ll}
		\vec{k}_0 = \frac{\pi}{a}(\phantom{-}1,\phantom{-}1,\phantom{-}1) & \vec{k}_1 = \frac{\pi}{a} (-1,-1,\phantom{-}1) \:, \\
		\vec{k}_2 = \frac{\pi}{a}(\phantom{-}1,-1,-1) & \vec{k}_3 = \frac{\pi}{a}(-1,\phantom{-}1,-1) \:,
	\end{array}
\end{equation}
where $a$ is the lattice constant. In idealized NWs with translation period $\vec{T}=a(1,1,0)/2$ two of the valleys, $L_0$ and $L_1$, project on the NW axis to the edge of the NW's Brillouin zone ($k=\pi/T$), while the others, $L_2$ and $L_3$, to the $\Gamma$ point ($k=0$). 
Therefore $L_0\to L_1$ and $L_2\to L_3$ intervalley scattering processes are independent.
In each of them, the mass anisotropy does not contribute to the valley splitting, since $m^*_0=m^*_1$ and $m^*_2=m^*_3$ with respect to the NW axis.
Following~\cite{Poddubny12} one may estimate the matrix element of the intervalley scattering at the NW surface as
\begin{equation}
	\label{eq:M_sum}
	M_{\mu, \nu} = C \sum_n e^{- i (\vec{k}_{\mu} - \vec{k}_{\nu}) \vec{R}_n} \Phi^*_{\mu}(\vec{R}_n) \Phi_{\nu}(\vec{R}_n) \:,
\end{equation}
where $\Phi$ is an envelope function, $\vec{R}_n$ is a lattice node and $C$ is a microscopic constant, an intergal over the primitive cell. 
Despite the deceptive simplicity, evaluation of the matrix element is difficult as one has to compute the microscopic constant, exact envelopes and deal with the spin degeneracy. 
The spin can not be excluded due to its strong influence on the PbSe bandstructure. 
Eq.~\eqref{eq:M_sum}, however, helps to compare the two $L_0\to L_1$ and $L_2\to L_3$ scatterings processes.
One can show, that the phase factors, $e^{-i (\vec{k}_0-\vec{k}_1 )\vec{R}_n}$ and $e^{-i (\vec{k}_2-\vec{k}_3)\vec{R}_n}$, are the same, which assumes similar dependencies of the valley splitting on the microscopic NW structure.

\begin{figure}[t]
	\includegraphics{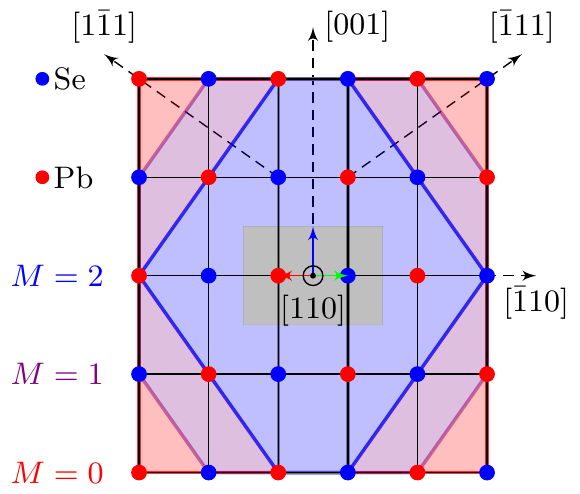}
	\caption{[110] view onto the NW supercells of type II with size paramter $N=2$ and different shape parameters $M$. The supercells with $M=0, 1, 2$ are enclosed in red (rectangular), violet and blue (rhombic) regions. The gray region in the middle corresponds to the smallest possible supercell of type II, shown also in Table~\ref{tab:NW_types}. The small RGB arrows show the $xyz$ tcrystallographic axes.}
	\label{fig:NM_shape}
\end{figure}

For numerical simulations we use the supercell approach~\cite{Cardona_book} without accounting for surface reconstruction and relaxation.
We control the structure of NW supercells by four integers: size parameter $N > 0$, shape parameter $M$ $(0 \le M \le N)$, and two additional numbers $dN_1, dN_2 \in \{0,1\}$ to adjust the NW symmetry.
The NW supercells are prepared in two steps: first we make a parallelepiped consisting of $(2N+2+dN_1) \times (2N+1+dN_2)$ atomic layers along the $[\bar110]$ and $[001]$ directions. 
Due to the periodicity, the supercell is only 2 atomic layers thick along the translation vector $\vec{T}$.
Then we form the NW shape by cutting $M$ $\{111\}$ atomic layers from the corners of the parallelepiped along $[1\bar11]$ and $[\bar111]$ directions, as shown in Fig.~\ref{fig:NM_shape}.
Below we will refer the NWs with $M=0$ as ``rectangular'' and the ones with $M=N$ as ``rhombic''.

In total we consider four different NW types with $D_{2h}$ (I), $C_{2v}$ (II, III) and $C_{2h}$ (IV) symmetries, summarized in Table~\ref{tab:NW_types}.
Note that NWs of type II and III have different orientation of $C_2$ axes and the NWs of type IV have non symmorphic spatial group. 
More details on the NW symmetry is given in Appendix~\ref{app:geom}.

\begin{ruledtabular}
	\begin{table}[b]
		\caption{Four considered NW types and their smallest supercells ($N=0$). 
		The cation, located at $(0,0,0)$ in the bulk coordinates frame, is marked with ``x''.}
		\begin{tabular}{ccccc}
			&
			\includegraphics{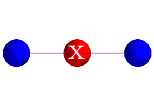}
			& 
			\includegraphics{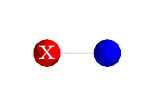}
			& 
			\includegraphics{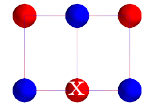}
			&
			\includegraphics{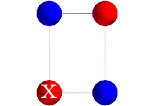}
			\\ \hline
			NW Type & I & II & III & IV \\ \hline
			symmetry & $D_{2h}$ & $C_{2v}$ & $C_{2v}$ & $C_{2h}$ \\ \hline
			$dN_1, dN_2$ & 1, 0 & 0, 0 & 1, 1 & 0, 1 \\
		\end{tabular}		
		\label{tab:NW_types}
	\end{table}
\end{ruledtabular}

\section{Results and discussion}
In this work we focus on NWs of intermediate size, where the valley splittings are large enough and the NWs are not too small. 
It was shown~\cite{Avdeev17}, that the valley splitting is negligible (compared to exchange energy~\cite{Zunger07}) in relatively large systems and is the most pronounced for wires with diatemeters $\lesssim 50$ \AA.
Here we limit ourselves with size parameters $2 \le N \le 9$ corresponding to the range of NW cross section sizes from $\approx 10$ to $\approx 50$ \AA.

Regarding the symmetry, the first three types of NWs are simple.
Indeed, the only spinor representations of $D_{2h}$ (type I) and $C_{2v}$ (II, III) are two dimensional $\Gamma_5^{\pm}$ and $\Gamma_5$, respectively~\cite{Koster}.
In the $D_{2h}$ NWs, with inversion center being at cation, consecutive valley multiplets have opposite parity, therefore the valley multiplets are split via self-scattering at the NW surface.
In the $C_{2v}$ NWs, the lack of spatial inversion also allows for the valley coupling with high energy multiplets, though this assumed to be negligible.
In the $C_{2h}$ NWs (type IV), the spatial symmetry forbids the splitting of the $L_0, L_1$ multiplets and the main mechanism of the $L_2, L_3$ valley splitting becomes the coupling with high energy states.
Detailed analysis of this case is given below and in Appendix~\ref{app:symm}.

\begin{figure}[t]
	\includegraphics{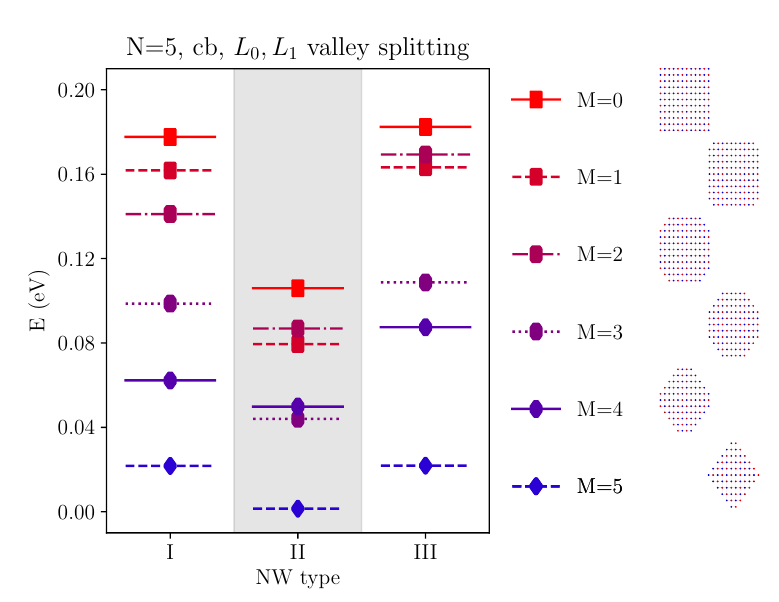}
	\caption{Valley splitting energies of the $L_0, L_1$ ground conduction multiplet in NWs of type I, II and III as a function of shape parameter $M$. Corresponding NW supercells (of type II) are shown on the right. The size parameter $N=5$.}
	\label{fig:T_N5_L01_vs}
\end{figure}

In Fig.~\ref{fig:T_N5_L01_vs} we show the valley splitting energies of the $L_0, L_1$ ground conduction multiplet as a function of the shape parameter $M$ in NWs of the first three types (see Table~\ref{tab:NW_types}) with the shape parameter $N=5$ (approximate lateral size $25\times30$ \AA).
The valley splittings are maximal in rectangular NWs ($M=0$) and almost absent in rhombic NWs ($M=N$).
These NWs have their surfaces terminated either by $\{001\}$ and $\{110\}$ facets, or mostly by $\{111\}$ facets, respectively.
The very similar dependence on the parameter $M$ holds for each $N$.
Even thought the $\{111\}$ facets of PbSe are polar, in this work we assume that this charge is compensated and therefore neglect the built-in electric field.
Note that the NWs of type III have one extra $(\bar110)$ and one extra $(001)$ atomic planes compared to the NWs of type II, but their splitting energies of the ground $L_0, L_1$ multiplet are almost twice as large.
The type IV is not shown here, as in this case the $L_0, L_1$ valley multiplet acquires an extra degeneracy due to the time reversal symmetry, see Appendix~\ref{app:symm}.

\begin{figure}[t]
	\includegraphics{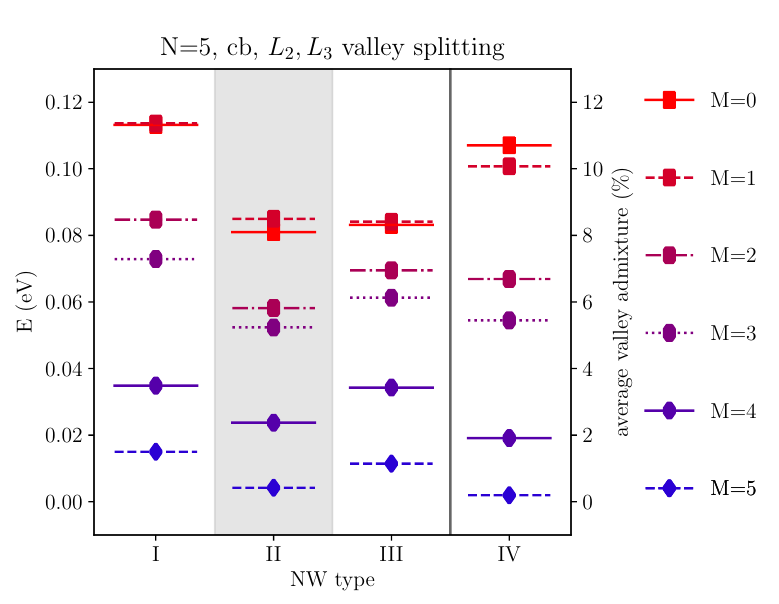}
	\caption{NW types I, II, III: same as in Fig.~\ref{fig:T_N5_L01_vs}, but for the $L_2, L_3$ ground conduction multiplet. 
	NW type IV: relative average valley admixture, $(\Delta \rho_2 + \Delta \rho_3)/2$, as a function of shape parameter $M$. 
	The size parameter $N=5$.}
	\label{fig:T_N5_L23_vs}
\end{figure}

The valley splitting energies of the $L_2, L_3$ ground conduction multiplet as a function of the shape parameter $M$ in NWs with $N=5$ are summarized in Fig.~\ref{fig:T_N5_L23_vs}.
They are only about two thirds of the corresponding splittings of the $L_0, L_1$ multiplet, which is mostly due to the mass anisotropy.
Indeed, the $L_2$ and $L_3$ valleys lie within (110) plane, while the $L_0$ and $L_1$ are tilted towards the wire axis and therefore have lighter effective in-plane masses, see~\eqref{eq:mass}.
Also note, that there is almost no difference in splitting energies in the NWs of type II and III.

Now let us turn to the NWs of type IV, shown in right column in Fig.~\ref{fig:T_N5_L23_vs}. 
In these NWs the $L_2$ and $L_3$ ground valley multiplet is split via completely different mechanism.
Since inversion center in NWs of type IV is located between atoms (see Table.~\ref{tab:NW_types}), the $L_2$ and $L_3$ valleys have different parity and can not mix directly.
The parity of an $L_{\mu}$ valley is related to phase factors $e^{i \vec{k}_{\mu} \vec{R}}$ at lattice nodes. 
For $\vec{R}=\pm a(0,1,1)/4$, positions of the two closest to inversion center cations, these phases are $\mp i$ for the $L_2$ valley and $1$ for the $L_3$.
This implies that $L_2~(L_3)$ valley is odd (even) and that the valley coupling in this case is only possible via far energy states.

\begin{figure}[h]
	\includegraphics{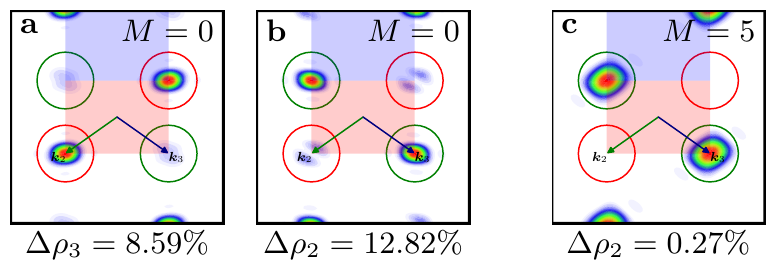}
	\caption{kLDOS of the $L_2, L_3$ ground conduction multiplet states in NWs of the type IV.	
	(a) Admixture of $L_2$ valley state to $L_3$ valley state in rectangular NW. (b) Admixture of $L_3$ to $L_2$ in rectangular NW, (c) same in rhombic NW.
	Red (green) circles show the area used to estimate the local density $\rho_{2(3)}$ near the $L_{2(3)}$ valley.
	The size parameter $N=5$.
	}
	\label{fig:kLDOS}
\end{figure}

In fact, this possibility is realized and is the most pronounced in rectangular NWs (shape parameter $M=0$), similarly to the valley splittings.
To illustrate this, in Fig.~\ref{fig:kLDOS} we show the local density of states in reciprocal space (kLDOS, see Refs.~\cite{Nestoklon16_Ge,Avdeev17} for details).
For quantitative description of the valley coupling we introduce ``valley admixture''
\begin{equation}
	\label{eq:vadm}
	\Delta\rho_{2(3)} = \frac{\rho_{2(3)}}{\rho_2 + \rho_3} \:,
\end{equation}
where $\rho_{2(3)}$ is the valley density, integrated over small region near the $L_2~(L_3)$ valley shown by red (green) circles in Fig.~\ref{fig:kLDOS}. 
One can see that the valley admixture in rectangular NWs is much higher that in rhombic ones. 
The value $\Delta\rho_3$ in rhombic NW ($M=N=5$) is $0.11$\% and the corresponding kLDOS is not shown in Fig.~\ref{fig:kLDOS}.
The density distributions near the main peaks in Figs.~\ref{fig:kLDOS}a--\ref{fig:kLDOS}c are $s$-like (correspond to the ground multiplet), while distributions near the secondary peaks in Figs.~\ref{fig:kLDOS}a and \ref{fig:kLDOS}b are $d$- and $p$-like, respectively.
It proves that the admixed states belong to different valley multiplets.
Note, for convenience in Fig.~\ref{fig:T_N5_L23_vs}~we show the average valley admixtures $(\Delta \rho_2 + \Delta \rho_3)/2$ for the NWs of type IV.

\begin{figure}[t]
	\includegraphics{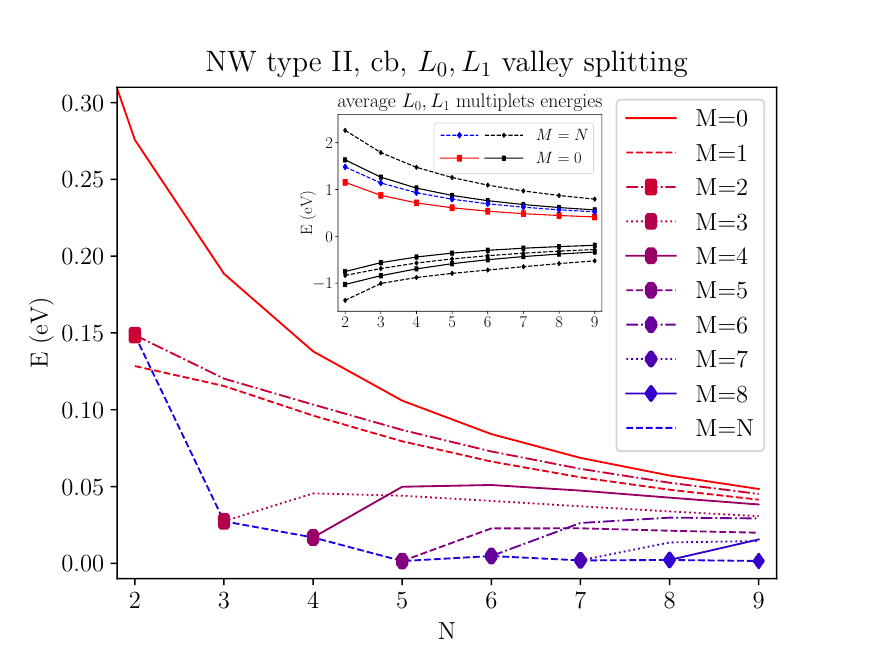}
	\caption{The valley splitting energies of the $L_0, L_1$ ground conduction multiplet in NWs of type II as a funciton of the NW shape ($M$) and size ($N$) parameters.
	Lines connect data point with the same $M$. Each line starts from $N=M$ (indicated by marker) and has a unique combination of color and style.
	Inset: average energies of the two lowest energy valley multiplets in conduction and valence bands in rhombic ($M=N$, solid) and rectangular ($M=0$, dashed) NWs as a function of their size parameter $N$.	}
	\label{fig:T2_N_vs_L01}
\end{figure}

\begin{figure}[t]
	\includegraphics{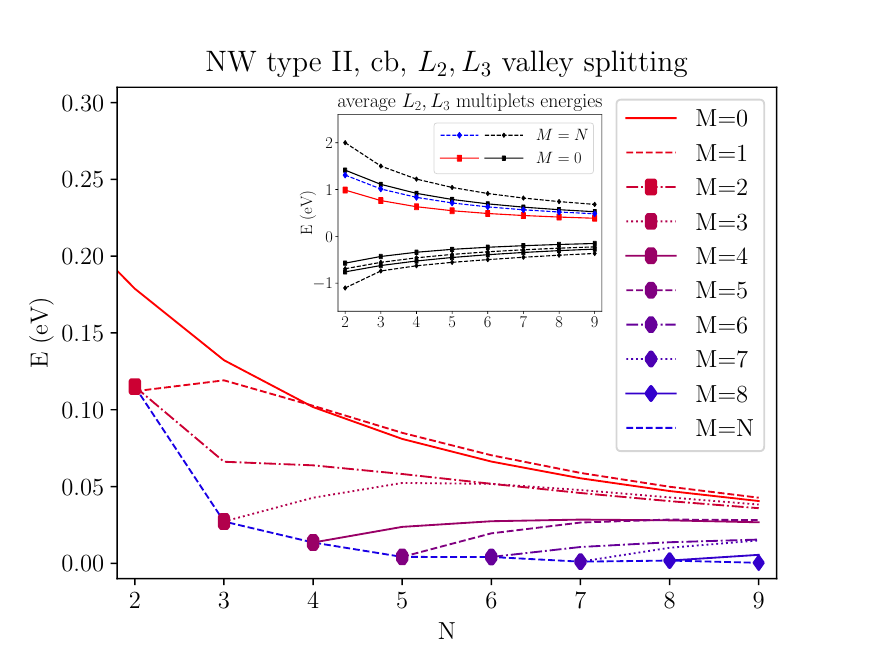}
	\caption{Same as in Fig.~\ref{fig:T2_N_vs_L01}, but for $L_2, L_3$ multiplets.}
	\label{fig:T2_N_vs_L23}
\end{figure}

Before we discussed NWs with the size parameter $N=5$ (ten atomic layers along the $[001]$ and $[\bar110]$ directions).
Next, in Figs.~\ref{fig:T2_N_vs_L01} and~\ref{fig:T2_N_vs_L23} we show both size ($N$) and shape ($M$) dependencies of the valley splitting. 
We show the data only for the NWs of type II, since the others behave very similar.
We vary the size parameter $N$ from $2$ to $9$ so the lateral size of NWs changes from
$11\times 12$ \AA$^2$ to $41 \times 55$ \AA$^2$ along the $[\bar110]$ and $[001]$ axes respectively.
In order to highlight the shape and size dependencies, in Figs.~\ref{fig:T2_N_vs_L01} and~\ref{fig:T2_N_vs_L23} we connect data points with the same parameter $M$ by color lines (solid red for $M=0$, dashed blue for $M=N$). 
For each size parameter $N$, the shape parameter $M$ satisfies $0\le M\le N$, therefore each line in Figs.~\ref{fig:T2_N_vs_L01} and~\ref{fig:T2_N_vs_L23} starts from $N=M$. 
The starting point of each line (except for $M=0$ and $M=N$) is indicated by marker.

Fig.~\ref{fig:T2_N_vs_L01} shows the valley splitting energies of the $L_0, L_1$ ground conduction multiplet. 
Inset shows the average energies (without the valley splitting) of the two valley multiplets with lowest energies in conduction and valence bands in rectangular ($M=0$, dashed lines) and rhombic ($M=N$, solid lines) NWs as a function of size parameter $N$.
Average energies of the ground conduction multiplet are additionally indicated by color to match the main plot (red for $M=0$, blue for $M=N$). 
Note, that the rhombic NWs exhibit larger confinement energies than their rectangular counterparts, while the valley mixing in rhombic NWs is strongly suppressed. 

The same data for the $L_2, L_3$ ground conduction multiplet is shown in Fig.~\ref{fig:T2_N_vs_L23}.
We do not discuss it in details, but mention, that all the size and shape dependencies of the valley splitting (coupling) are almost the same, except for their absolute values.
Therefore we conclude, that the shape induced suppression of the valley splitting in NWs is rather a physical phenomena.

\section{Conclusion}
We studied the valley splitting in $[110]$-grown nanowires with different size, shape and symmetry.
We demonstrate, that the valley splittings substantially depend on the NW shape, for particular case of prismatic octagonal NWs which is determined by relative fraction of $\{001\}, \{110\}$ facets compared to $\{111\}$ facets at the NW surface.
The values of valley splittings are large, up to $100$ meV in NWs about $5$ nm diameter. 
The splittings tend to have maximal values in rectangular NWs ($\{001\}$ and $\{110\}$ facets at the surface) and are almost absent in rhombic NWs (mostly $\{111\}$ facets at the surface).
This result holds for a wide range of NW sizes and different point symmetries.

We also found a special type of NWs with non symmorphic spatial group, where $L_0, L_1$ valley multiplets become fourfold degenerate and the splitting of $L_2, L_3$ multiplets is due to the intervalley coupling via far energy states

Results of this work, except for the absolute values of the valley splittings, also apply for PbS and PbTe, due to the very similar band structures of these materials.

\begin{acknowledgments}
The author acknowledges fruitful discussions with M.~O.~Nestoklon and financial support by RFBR Project No. 17-02-00383 A.
\end{acknowledgments}

\appendix
\section{Microscopic wire structure}
\label{app:geom}

PbSe has the rocksalt crystal structure with
\begin{equation}
	\label{eq:lvs} 
	\vec{a}_1 = \frac{a}2(1,0,1) \:,
	\quad
	\vec{a}_2 = \frac{a}2(1,1,0) \:,
	\quad
	\vec{a}_3 = \frac{a}2(0,1,1) \:
\end{equation}
lattice vectors, 
where $a=6.1$ \AA\ is the lattice constant~\cite{Poddubny12}. Reciprocal lattice vectors are conveniently related with the $L$ valleys~\ref{eq:kks} as
$\vec{b}_{\mu} = 2 \vec{k}_{\mu}, \mu = 1, 2, 3$.

With the used tight binding parametrization~\cite{Poddubny12} the first conduction band effective mass ratios along the $[\bar110], [001]$ and $[110]$ directions ($xyz$ NW axes) are
\begin{equation}
	\label{eq:mass} 
	\frac{m_{0(1)}^{[\bar110]}}{m_{2(3)}^{[\bar110]}} = 0.74 
	\:,\qquad
	\frac{m_{0(1)}^{[001]}}{m_{2(3)}^{[001]}} = 1
	\:,\qquad
	\frac{m_{0(1)}^{[110]}}{m_{2(3)}^{[110]}} = 1.36 \:,
\end{equation}
where $m_{\mu(\nu)}^{\vec{n}}$ denotes the effective mass along $\vec{n}$ of the first conduction band electron at the $L_{\mu(\nu)}$ valley.

The first three NW types, see Table~\ref{tab:NW_types}, have $D_{2h}$, $C_{2v}$ and $C_{2v}$ point groups, respectively. 
Their symmetries are determined by orientation of $C_2$ axes and the position of the point symmetry origin relative to the $(0,0,0)$ cation, marked by ``x'' in Table~\ref{tab:NW_types}. 
NW type I has inversion center, three $C_2^x, C_2^y, C_2^z$ rotation axes (in the NW coordinates frame) and three corresponding reflection planes $\sigma_v^x, \sigma_v^y, \sigma_h^z$. 
NW type II has $C_2^x$ rotation axis and $\sigma_v^y, \sigma_v^z$ reflection planes with the point symmetry origin being at $a(-1,1,0)/8$. 
NW type III has $C_2^y$ rotation axis and $\sigma_v^x, \sigma_v^z$ reflection planes with the point symmetry origin being at $a(0,0,1)/4$.

NW type IV has non symmorphic spatial group with $C_{2h}$ point symmetry. 
We use $a(-1,1,2)/8$ point (relative to the $(0,0,0)$ cation) as the point symmetry origin, which results in the following quotient group
\begin{equation}
	\label{eq:choice2}
	\left\{e, a=\left(C_2^z, \frac{\vec{a}_3}2\right), b=\left(\sigma_h^{z}, \vec{0}\right), ab=\left(i, \frac{\vec{a}_3}2\right) \right\} \:.
\end{equation}
Note, that this is not the only possible, but is the most convenient way to choose the point symmetry origin.

\section{Symmetry analysis}
\label{app:symm}
Symmetry analysis for the first three NW types is trivial and therefore omitted. 
Instead, we focus on the NW type IV to describe the absence of the valley splitting in $L_0, L_1$ valley multiplets.

Following ref.~\cite{BirPikus}, we use projective representations to classify the states of $L_0, L_1$ valley multiplets, because these valleys project onto the edge of the NW Brillouin zone.
In $C_{2h}$ there are two projective classes: $K_0$ and $K_1$. 
Class $K_0$ has only one dimensional (vector and spinor) representations~\cite{Koster}, while in $K_1$ there is only one two-dimensional representation $P^{(1)}$.

Since the $L_2, L_3$ valleys project onto the $\Gamma$ point, states of the $L_2, L_3$ multiplets belong to the class $K_0$ and can be classified according to ref.~\cite{Koster}.
States of the $L_0, L_1$ valley multiplets belong to the class $K_1$, and therefore transform according to $P^{(1)}$.
Indeed, the factor system
\begin{equation}
	\label{eq:fs}
	\omega_{\vec{k}}(g_1, g_2) = e^{i (\vec{k} - R_1^{-1} \vec{k}) \vec{\tau}_2} \:
\end{equation}
on elements~\eqref{eq:choice2} with $\vec{k}=\vec{k}_0$ or $\vec{k}_1$,~\eqref{eq:kks}, has the form
\begin{equation}
	\label{eq:fs_k01}
	\omega_{\vec{k}_{0(1)}}(a^k b^p, a^{k'} b^{p'}) = \alpha^{(p k')} \qquad \alpha=-1\:,
\end{equation}
which is the standard form for the class $K_1$~\cite{BirPikus}. 
Note, since all $\vec{\tau}\parallel [110]$, the factor system depends only on projection of the wave vector $\vec{k}$ onto the NW axis.

Next we consider time reversal symmetry by means of Herring criterion~\cite{Herring37,BirPikus}, which is about relation of $\psi$ and $\hat{T}\psi$, where $\hat{T}$ is the time reversal operator. 
The criterion reads as a sum over quotient group
\begin{equation}
	\label{eq:H_cr}
	\frac1{h} \sum_{g\in G/T} \chi(g^2) = 
	\left\{
	\begin{array}{cc}
		K^2\:, & (a) \\
		0\:,  & (b) \\
		-K^2\:, & (c)
	\end{array}
	\right. 
	\:,\quad
	\hat{T}^2 = K^2 \hat{I}
	\:.
\end{equation}
There are three options: $\psi$ and $\hat{T}\psi$ are (a) llinearly dependent, (b) linearly independent and tranform according to conjugate representations or (c) equivalent representations. 
In cases (b) and (c) time reversal symmetry leads to additional degeneracy of states.

The fact that the factor system for $L_0,L_1$ multiplets, eq.~\eqref{eq:fs_k01}, has the standard form, allow us to use the explicit form of generator matrices $a=\sigma_z, b=\sigma_x$ for $P^{(1)}$ representation from ref.~\cite{BirPikus}. 
With generators matrices the sum~\eqref{eq:H_cr} for $L_0, L_1$ states is easily evaluated
\begin{equation}
	\frac{2 + \Tr(\sigma_z^2)+\Tr(\sigma_x^2)-\Tr(\sigma_y^2)}4=1 = -K^2 
\end{equation}
and we see that the case (c) is realized. 
Here $K^2=-1$ due to the antiunitary nature of time reversal operator $\hat{T}$ for spinors.
Therefore the states of $L_0, L_1$ multiplets are fourfold degenerate.

Note, that for $L_2, L_3$ states, the case $(b)$ is realized, so all the states of $L_2,L_3$ multiplets are twice degenerate. 
We also note, that evaluation of the sum~\eqref{eq:H_cr} in this case is much simpler using the double group approach~\cite{Koster}, instead of projective representations theory.

\bibliography{PbSe}
\end{document}